# A Service-Centric Approach to a Parameterized RBAC Service

JONATHAN KEIRRE ADAMS
Graduate School of Computer and Information Sciences
Nova Southeastern University
3301 College Avenue, Ft. Lauderdale, FL
UNITED STATES
jonaadam@nova.edu,
http://www.scis.nova.edu/~jonaadam

*Abstract:* - Significant research has been done in the area of Role Based Access Control [RBAC]. Within this research there has been a thread of work focusing on adding parameters to the role and permissions within RBAC. The primary benefit of parameter support in RBAC comes in the form of a significant increase in specificity in how permissions may be granted. This paper focuses on implementing a parameterized implementation based heavily upon existing standards.

## 1 Introduction

The benefits inherent in role based access control and its derived access control models make this paradigm very well suited for use in web-based applications. As such, this work sets out to create an extensible access control service for web services. There are several standards and a body of existing research in this vein, and this work will seek to build upon prior successes as much as possible, while extending the body of work as necessary.

There are many benefits inherent in having a flexible access control model that is capable of integrating different factors, for example, contextual information, into its access control policies. The addition of parameter support on top of the role based access control paradigm is an effective means of reaching this goal. Previous work has shown the benefit of utilizing contextual information as well as the overall benefits derived from parameterized RBAC. An access control service that uses parameterized role based access control to implement flexibility and extensibility in its policy engine will have a variety of applications. The use of parameterized roles allows a greater degree of specificity in terms of the access granted to a specific role. This will reduce the overall amount of roles, by allowing roles to grant different degrees of access based upon their parameterized input.

## 2 Background Information

The problem posed by this paper is the creation and implementation of an extensible access control service. To solve this problem, we look at the existing body of work and find Service-Oriented Role Based Access Control [SRBAC], a model derived from RBAC with the intention of providing access control for web services. Extensible Access Control Markup Language [XACML] is a critical standard for expressing access control policies, requests, and responses. XACML provides an excellent means of expressing RBAC policies in an interchangeable format.

This paper will investigate the feasibility of implementing an access control service based upon a variety of existing technologies that is reasonably flexible and extensible. The primary focus of the work is upon the margining of SRBAC and XACML. The benefit of having an RBAC approach to Service Oriented Access Control is work will be capable of synchronizing access control policy among a variety of related federated information sources. Utilizing XACML to implement SRBAC ensures that it will be adaptable to a variety of computing environments

Several researchers have explored bringing the functionality of Role Based Access Control to Web Services. Their work has taken many forms and has had differing approaches. In [18], Liu and Chen craft a model, based upon extensions to RBAC, called Web Services Role Based Access Control [WS-RBAC]. This work extends RBAC in taking into account the Business Process as a context to how a user attempts to access an object. Services are treated as objects, and access is largely governed via Business Processes, which serve as policies, marked up in Business Process Execution Language for Web Services [BPEL4WS]. BPEL4WS is a XML based markup language that is utilized for adding support for business related web services interaction [19]. Murty and Taylor in [20] devised a framework for controlling access to web resources utilizing Role Based Access Control and an XML format for passing access control information. Their work is primarily focused on distributing the responsibility for managing the access control database. Their work is now somewhat dated due to the fact that subsequent XML standards, such as XACML have arisen, yet their work takes a still-relevant approach to managing access control information, such as users, objects, etc.

### 2.1 Role Based Access Control

Role-Based Access Control is a flexible access control model, primarily notable for its addition of the notion of roles into access control. Ferriaolo and Kuhn have said that RBAC is a Mandatory Access Control [MAC] model, but that it diverges from MAC in several ways [3]. Many models have been derived from RBAC to solve a variety of access control related issues. Role Based Access Control has evolved significantly over the past decade, gaining acceptance. The notion of roles in RBAC provides significant benefit to managing access control in enterprise environments. RBAC allows the management of privileges by roles instead of subjects and supports the principle of least privilege and separation of duties [11].

A significant benefit to the underlying paradigm of RBAC is its adaptability to various computing configurations and constraints used to determine access control decisions. Researchers have adapted RBAC for geo-spatial applications, for ubiquitous computing applications, as well as extending aspects of the model to support features such as cascading delegation, negative permissions, and a variety of others that make the model more flexible and/or easier to manage in an actual implementation. As the uses for computing and data increase in diversity, access control models have been adapted, significantly increasing in complexity. This is particularly true in pervasive computing and distributed computing applications. Because of the nature of these computing applications, many assumptions used in crafting an access control model must be reconsidered. A significant part of the added complexity comes in the form of considering other factors. In the case of this work, the goal is to find a suitable extension of RBAC that is adaptable to the constraints involved in web services, while adding the benefits derived from a parameter based approach to specifying roles.

### 2.2 Parameterization

In [2], parameterization is adding descriptive information to roles and privileges to allow greater flexibility. In terms of privileges, [2] describes parameterized privileges as 2-tuples, where the first parameter represents the object to be accessed and the second parameter represents the access mode for that object given the privilege. With an object *customerData*, and permissions read, write, and null, example privileges would look like:

- $Priv_1$ = (customerData, read)
- $Priv_2$ = (customerData, write)
- $Priv_3$ = (customerData, null)

Thus, privileges are of the form $Priv_n = (o, p_n)$, where $p$ is a set of privileges for an object $o$. Parameterization also affects the subject portion of the RBAC equation. In [2], the authors describe roles with the notation *(rname, rpset, rparamset)* where *rparamset* is the set of role parameters for a given role with the name *rname*. The *rparamset* can be a null value if the role has no given parameters. An example role *customerDataService* that has an identifying attribute *className* with possible values *updateCustData, readCustData* and, *eraseCustData* could be represented as:

- *(customerDataService, className, updateCustData)*
- *(customerDataService, className, readCustData)*
- *(customerDataService, className, eraseCustData)*

In [2], it is notable that *PA*, the privilege to role assignment, is represented using XML XPath notation. Because of this, a privilege to role assignment example incorporating the above-explained notation could be represented as:

- $pa_1$ = (//customerDataService[@className = updateCustData]/customerDataService,update)
- $pa_2$ = (//customerDataService[@className = updateCustData]/customerDataService,read)
- $pa_3$ = (//customerDataService[@className = updateCustData]/customerDataService,erase)

For this work, since it will be using XACML, the approach will differ. XACML links *PolicySets* using the *<PolicySetIdReference>* tag. In XACML 1.x, RBAC is implemented using linked *PolicySets* in this manner. Allowing for parameterization will simply require more complicated links between

*PolicySets*. This is represented by examples in the Fig. 1 and Fig. 2.

```xml
<PolicySet xmlns="urn:oasis:names:tc:xacml:1.0:policy"
  PolicySetId="RPS:student:role:studentid-02123781"
  PolicyCombiningAlgId=
  "urn:oasis:names:tc:xacml:1.0:policy-combining-algorithm:permit-overrides">
 <Target>
  <Subjects>
   <Subject>
    <SubjectMatch MatchId=
     "urn:oasis:names:tc:xacml:1.0:function:anyURI-equal">
     <AttributeValue DataType=
        "http://www.w3.org/2001/XMLSchema#anyURI">
         urn:example:role-values:student:rparams:studentid-02123781
        </AttributeValue>
     <SubjectAttributeDesignator
       AttributeId="urn:oasis:names:tc:xacml:1.0:subject:role"
       DataType="http://www.w3.org/2001/XMLSchema#anyURI"/>
    </SubjectMatch>
    <SubjectMatch MatchId=
       "urn:oasis:names:tc:xacml:1.0:function:string-equal">
     <AttributeValue
       DataType="http://www.w3.org/2001/XMLSchema#string">
                studentid-02123781
            </AttributeValue>
     <SubjectAttributeDesignator
       AttributeId="RParams"
       DataType="http://www.w3.org/2001/XMLSchema#string"/>
    </SubjectMatch>
   </Subject>
  </Subjects>
 </Target>
 <PolicySetIdReference>PPS:student:role:studentid-02123781
        </PolicySetIdReference>
</PolicySet>
```

Fig. 1 *XACML Representation of Parameterized Roles*

```xml
<Actions>
  <Action>
    <ActionMatch
         MatchId="urn:oasis:names:tc:xacml:1.0:function:string-equal">
      <AttributeValue
 DataType="http://www.w3.org/2001/XMLSchema#string">
                register</AttributeValue>
      <ActionAttributeDesignator
 AttributeId="urn:oasis:names:tc:xacml:1.0:action:action-id"
 DataType="http://www.w3.org/2001/XMLSchema#string"/>
      <AttributeValue
 DataType="http://www.w3.org/2001/XMLSchema#string">
                studentid-02123781</AttributeValue>
      <ActionAttributeDesignator
 AttributeId="AParams"
 DataType="http://www.w3.org/2001/XMLSchema#string"/>
    </ActionMatch>
  </Action>
</Actions>
```

Fig. 2 *XACML Representation of a Parameterized Privilege Instance*

## 2.3 Service Oriented RBAC

Service Oriented Role Based Access Control is a derivative of Role Based Access Control that is designed to account for basic RBAC unsuitability [5] as an access control model for a web services framework. A notable aspect of SRBAC is the presence of a function mapping for actors to single users. In SRBAC, the actor object is designed as a role activation proxy for a user's roles [5]. Also of note is the many-to-many service relation *SR* [5]. In SRBAC, services replace the object construct found in standard RBAC. Thus, the *SR* relation replaces the object relation found in RBAC. SRBAC is a simple and elegant solution to the modification of RBAC to facilitate controlling access in a service-oriented environment, such as that of web services.

### 2.3.1 Reasons for adding parameterization to SRBAC

Ge and Osborn's motivation for role parameterization comes into play for XML documents where a greater degree of specificity in determining which data is accessible to specific users, or user roles, than is possible with standard RBAC. In terms of web services, the same model may apply, as these services provide federated access to data sources and applications that may need to be constrained in a manner similar to the XML data sets that Ge and Osborn discuss. This work will focus on providing an integrated access control framework for controlling access to multiple related web services.

An example environment where an integrated access control framework with the flexibility provided by parameterized roles would be usable is a federated data source, such as a service that provides a dynamic XML or RSS feed of a news source. Parameterized roles could be used as a means to customize the content of such a data source based upon client attributes. Such work has the potential for the incorporation of such contextual information as the geographic location of the client or type of client, i.e. provide one set of content to an end user with a feed-reader versus a different content set to a partnering website based upon the user agent environment variable, which would be passed as a role parameter. Implementing an integrated role based access control system with parameterized roles could be beneficial in tightly filtering visible information to that absolutely required to be returned to the end user, based upon their needs as determined by their role within the framework.

### 2.3.2 Implementing SRBAC with XACML

The benefit of adding parameterized roles and privileges to SRBAC would come in the way Actors are activated. In Ge and Osborn's [2] paper, they use the construct of Actors as special proxy roles that are a part of Simple Object Access Protocol [SOAP] messages. Actors are activated for each role owned by a particular user. In this work, we chose not to use SOAP, thus adding parameter support to Roles will give us the same net effect, but in a way that can fit into the existing XACML format with minimal modification required. The *Actor* in SRBAC consists of <*user, role, time, constraints*>, it is possible for this work to represent this object using parameters. This could be done by creating dynamic roles with parameters represents a user identifying value, time, and any constraint or constraints required by a given system of services.

In [5], the authors chose to utilize SOAP messages as the medium for sending access control requests and responses, making the proxy role *Actor* necessary. However, because this work has utilized XACML instead of SOAP, the *Actor* is not necessary. In order to be true to the model, this work describes two approaches, one utilizing the *Actor*, and another, which does not require it. Essentially, the *Actor* is a pseudo-role. Implementing such a pseudo role in XACML is does not seem to be directly supported, but is possible by utilizing multiple requests to generate the credentials necessary to grant the requested access.

It is notable that, any implementation of the *Actor* as described in [5] will require specific application code to support the proxying of requests. For example, if a student wants to register for classes, and the system implements parameterized roles, the application code that approves the first request will activate the student's role and generate a hash or checksum to validate the request. The application code will submit a second request using the user's activated role as a parameter, that will activate the *Actor*, then the access to the object will be allowed. It is important that this second request have some verifying information, in our case a hash that is partially generated by some secret present on the server, to preserve the system's integrity.

Without the *Actor*, it is possible to submit the request without the secondary request. The second request does provide a degree of security to the system, in that the user has already been authenticated, and that the privilege is not directly associated with the user's role. The rationale given in [5] for having the second request is largely based upon the reliance on SOAP, which is not a integral part of this work. The choice to utilize the *Actor* is largely a design decision based upon the implementation needs of the system, balancing additional security against the additional overhead of addition request/response traffic. For the purpose of this work, we will not utilize the *Actor*.

## 2.4 Service Oriented RBAC

Extensible Access Control Markup Language is a standard of the Organization for the Advancement of Structured Information standards [OASIS] as of 2003. The standard is currently in its second revision as of February 2005. The goal of this standard is to provide a flexible, platform independent, environment independent means for specifying access control policy for XML content sharing. This open standard approach has been implemented by a variety of vendors, to include Sun Microsystems and the Java 2 Platform, as well as Microsoft's .NET environment. The quick acceptance of the XACML standards by vendors greatly increases its viability as a means for specifying access control policies in the web services environment.

### 2.4.1 Support for RBAC in XACML

It is notable that as of the publication time of SRBAC, the XACML standard did not directly support the notion of roles, [5] which are essential in RBAC and RBAC derived access control models. In the meantime, support has been added for roles in the form of an XACML profile for role-based access control in 2004. This document is specifically designed to extend XACML such that it will fulfill the requirements necessary to support RBAC without changes to the 1.0 or 1.1 standard [7].

With the second release of the XACML specification, a new profile for role based access control has been specified. This profile includes the necessary support for roles, but also, importantly, specifically supports core RBAC, also called $RBAC_0$ and hierarchical RBAC, also called $RBAC_1$ [8]. Within the context of this work, adding parameters to roles does not seem to violate any aspect of $RBAC_0$ or $RBAC_1$, although in the implementation, there may issues involved in formatting parameterized roles such that they will fit into the XACML implementation framework as described in this work.

For the implementation phase of this project, the author chose Sun Microsystems' XACML implementation. This is the only widely available open source implementation of XACML. Sun's implementation is Java-based, which helps to make it the ideal choice for implementing a web-service or web service enabling library. Sun's XACML code also requires an external XML parser, due to some limitations in the parser (Crimson) that it is included in Java 1.4.x. Among its limitations, the Crimson parser does not support XPath.

### 2.4.2 Use of Parameterized Roles in XACML

Working with XACML version 1, the notion of roles in RBAC is best handled in a way that this paper will describe as *policy layering*. In Policy Layering, One XACML policy is used to determine which subjects belong to a role. Other XACML policies can be linked, and through them, privileges can be derived.

## 3   Problem Solution

In this work, the problem consists of several parts that will be dealt with independently. First is the implementation of parameterized roles in within a framework that utilizes XACML. This was expected to be a potential problem because RBAC that supports parameterized roles was initially implemented using SOAP messages as a medium. This work is specifically focused on using XACML, as its medium, primarily to facilitate portability into environments beyond web services. The problem is, because this work itself does not use SOAP, it is to some degree, deviating from what the prior research has done.

### 3.1 Integrating RBAC and Web Services

XACML and Security Assertion Markup Language [SAML] have been established as standards for communicating access control requests and decisions. SAML handles authentication requests and responses, while XACML is a markup language for representing access policy [13]. This work will depend heavily on XACML. While some researchers utilize XACML [5], others have taken alternate approaches, in some cases using SOAP [6]. In this work, the policy based interactions, policy exchange and access requests and responses are handled in XACML, but in a web-services environment, these messages would likely be encapsulated in SOAP messages. It is notable that this differs from the approach in Ge and Osborn in [5. The difference is that the access control attributes in their work are passed as attributes of the actual SOAP message, versus encapsulating the access control request within the message.

## 4. Conclusion

This work accomplishes three research goals. First, it proves that XACML can be utilized to accommodate the notion of parameterized role based access control. Secondly, XACML can be used to accommodate Service Oriented Role Based Access Control, as described in [5] with minor modification. Finally, it shows how the benefits from parameterized roles and a service oriented version of role based access control can be combined to produce a very flexible access control paradigm for controlling access to federated sources, such as web services, without a requirement for modifying SOAP messages. An access control system such as that described in this paper could be implemented in a variety of forms, as a middleware component, as a module in an application, or as a server side component in a client-server application. Because of its protocol independence, a system could benefit from this access control system utilizing a variety of communication frameworks, peer-to-peer, client-server, or otherwise.

To test the basic feasibility of interpreting requests we first run some basic tests with the XACML Policy Decision Point. We first use basic XACML formatted policies and requests to establish for our purposes how the Sun XACML implementation handles access requests. A policy is then crafted that implements both parameterized roles and parameterized privileges attempting to access service related objects and run several tests that confirm that the Sun XACML implementation responds as expected. Our tests worked as expected and serve as a basis that can be built upon in establishing a comprehensive access control infrastructure that uses parameterized roles and parameters as a basis for a flexible access control policy implementation.

### 4.1 Future Work

There are a few avenues for further research. First, the actual application of this work to different application environments, for example, web services, peer-to-peer environments, as well in hybrid configurations would be well served by further treatment. Secondly, building an access control mechanism based upon the work in this paper and incorporating the means to provide enterprise subject, role, service, and object management facilities is an obvious direction for follow-on work. Also, much of the work described in this paper, coupled with an as yet specified infrastructure could be applied to an enterprise digital rights management, studying the problems inherent in such an approach would be a possible area of study. Another potential area for work would be in devising a new RBAC model that keeps the benefits of parameterization of roles, yet is designed specifically with the requirements of being able to operate in a variety of distributed systems environments.